\documentclass[lettersize,conference]{IEEEtran}
\IEEEoverridecommandlockouts
\usepackage{cite}
\usepackage{amsmath,amssymb,amsfonts}
\usepackage{algorithmic}
\usepackage{graphicx}
\usepackage{textcomp}
\usepackage{xcolor}

\usepackage{booktabs, multirow} 
\usepackage{subfigure}
\usepackage{float}
\usepackage{makecell}

\usepackage{newtxtext}
\usepackage{pgfplots}
\usetikzlibrary{calc}
\pgfplotsset{compat=newest}

\pgfplotsset{
    customlegend/.code={

    },
}

\usepackage{tcolorbox}

\definecolor{lightgreen}{RGB}{144,238,144} 
\definecolor{plotRed}{RGB}{255,67,67}
\definecolor{plotBlue}{RGB}{65,105,225}
\definecolor{plotGreen}{RGB}{50,205,50}
\definecolor{plotPurple}{RGB}{147,112,219}
\definecolor{plotOrange}{RGB}{255,140,0}
\definecolor{plotYellow}{RGB}{255,255,0}

\def\BibTeX{{\rm B\kern-.05em{\sc i\kern-.025em b}\kern-.08em
    T\kern-.1667em\lower.7ex\hbox{E}\kern-.125emX}}

\begin{document}

\title{Benchmark-based Study of CPU/GPU Power-Related Features through JAX and TensorFlow}




\author{\IEEEauthorblockN{Roblex Nana Tchakoute\IEEEauthorrefmark{1},
Claude Tadonki\IEEEauthorrefmark{1}, Petr Dokladal\IEEEauthorrefmark{2} and Youssef Mesri\IEEEauthorrefmark{3}
}
\IEEEauthorblockA{\textit{Centre de Recherche en Informatique (CRI), Mines Paris - PSL University, Fontainebleau, France\IEEEauthorrefmark{1}}\\
\textit{Centre de Morphologie Mathématique (CMM), Mines Paris - PSL University, Fontainebleau, France\IEEEauthorrefmark{2}}\\
\textit{Centre de Mise en Forme de Matériaux (CEMEF), Mines Paris - PSL University, Sophia Antipolis, France\IEEEauthorrefmark{3}}
}
\IEEEauthorblockA{Email : \{roblex.nana\_tchakoute, claude.tadonki, petr.dokladal, youssef.mesri\}@minesparis.psl.eu}
}

\maketitle
\begin{abstract}
Power management has become a crucial focus in the modern computing landscape, considering that {\em energy} increasingly stands as a critical resource. This boosted the importance all topics related to {\em energy-aware computing}. This paper presents an experimental study of three prevalent power management techniques that are {\em power limitation, frequency limitation}, and {\em ACPI/P-State governor modes} (OS states related to power consumption). Through a benchmark approach with a set of six computing kernels, we investigate {\em power/performance} trade-off with various hardware units and software frameworks (mainly TensorFlow and JAX). Our experimental results show that {\em frequency limitation} is the most effective technique to improve {\em Energy-Delay Product (EDP)}, which is a convolution of energy and running time. We also observe that running at the highest frequency compared to a reduced one could lead to a reduction of up to factor $\frac{1}{10}$ in EDP. Another noticeable fact is that frequency management shows a consistent behavior with different CPUs, whereas opposite effects sometimes occur between TensorFlow (TF) and JAX with the same power management settings.
\end{abstract}


\section{Introduction}
There growing need for high-performance computing (HPC)  in various fields including  {\em artificial intelligence} and {\em scientific/technical simulations} at large scales \cite{tadonki2013high}. However, cutting-edge HPC infrastructures consume a noticeable amount of energy for their main purposes and maintenance-related aspects, thus an increasingly important focus on {\em power consumption} in the design and management of modern computing systems. AI power consumption is projected to rise significantly over the coming years, with particularly worrying predictions. Schneider Electric estimates that AI's power demand will grow from 4 GW (35 TWh annually) in 2023 to 15 GW (131 TWh annually) by 2028 \cite{se2024}. Other predictions indicate that AI workloads might consume between 85 TWh and 134 TWh by 2027, potentially increasing total energy demand of data centers by 30-50\% \cite{alex2023}. An annual consumption of 5.8 TWh (resp. 21.9 TWh) was expected from the use of Nvidia's AI infrastructure in 2024 (resp. 2025), which represents 7.3\% of global data center power consumption \cite{uptime2024}. These projections reflect AI’s growing energy consumption. 

As data centers and computing infrastructures scale up, optimizing for energy  contributes to the reduction of operational costs as well as associated environmental impacts \cite{nana2023energy}. Power management techniques typically target both processing performance and energy efficiency \cite{BENINI2002231}, particularly with heterogeneous computing systems. For example, Intel’s Icelake SP and AMD’s Zen3 CPUs, along with Nvidia’s A100 GPU, are high-end platforms that target high processing performance while offering various power management features. However, the effectiveness of those power management techniques is highly dependent on {\em hardware specificities}, {\em workload characteristics}, and {\em considered applications} \cite{Nadja2016}.

This paper focuses on evaluating the impact of three major CPU/GPU power management techniques that are {\em power limitation}, {\em frequency limitation}, and {\em ACPI/P-State governor modes}. We investigate the impact of power management techniques under different power-related settings and  directives through a benchmarking with six computing kernels that are \{compute, memory, mixed\}-bound.

In addition to native {\em hardware mechanisms}, {\em software features} play a crucial role in the management of power efficiency. TensorFlow (TF) \cite{tensorflow2015-whitepaper} and JAX \cite{jax2018github}, two widely used libraries for machine learning and numerical computing, are known for their distinct computational models and levels of performance. Our study is conducted through the comparison of these two frameworks under various power management configurations, thereby providing insights about how software features handle power-performance trade-offs.

The contributions of this paper are:
\begin{itemize}
    \item an empirical evaluation of three popular power management techniques on high-end processors.
    \item a differentiated study of TF and JAX under the same power management settings, showcasing how framework-specific optimizations can affect power efficiency.
    \item Platform and workload specific insights to guide practitioners in selecting appropriate energy management settings/strategies.
\end{itemize}

This remainder of the paper is organized as follows. Section \ref{sec:literature}  provides an overview of the related work on energy efficient computing. Section \ref{sec:experimental} describes the technical background and our experimental testbed for this study. Section \ref{sec:analysis} presents and comments our experimental results through power management settings with focus on EDP and power consumption, while Section \ref{sec:discuss} discusses the implications and limitations of our findings. Section \ref{sec:conclusion} concludes the paper and indicates some perspectives.

\section{Related work}\label{sec:literature}
We provide a overview of some relevant contributions on the main aspects of power-related research works.
\textbf{Processor efficiency:} Many studies have investigate energy through standard efficiency of processors. Cebrian et al. \cite{Cebrian2020} analyze the impact of vector computing (with (SSE, AVX, AVX512)) on {\em instructions throughput}, {\em reduction}, and {\em instructions per cycle (IPC)} and thereby demonstrate its potential for energy efficiency improvement. The Energy-Delay Product (EDP), a metric that correlates execution time and energy, was considered as a key efficiency metric in a work by Horowitz et al. \cite{Horowitz1994, Juan2012} and it so far remains pivotal in the trade-off analysis of energy-performance as in this work to. While performance-counter-based power modeling has been explored \cite{Dolz2016, Cupertino2015}, clearly identifying their impact on power management technique remains to be more investigated.

\textbf{Benchmarking \& power management:} Evaluating energy efficiency requires robust benchmarking across considered platforms. A comparative study by Suarez et al. \cite{Suarez2024} using various architectures suggest that device-specific features often play a dominant role in energy efficiency than  standard characteristics. Similar to our context, a work by Schöne et al. \cite{Schone2021} analyses power management mechanisms on AMD's Zen2 architectures and illustrates the impact of {\em frequency transitions} and so-called {\em P-states}.

\textbf{Energy of memory activity:} The energy cost of I/O w.r.t main memory is not negligible. A work by Schöne et al. \cite{Schone2012} investigates memory system power, while Catal\'{a}n et al. \cite{Cataln2014AnalyzingTE} explore energy overheads across the memory hierarchy.  The impact of workload size and memory usage on power scaling on multi-core CPUs using linear algebra kernels is illustrated by Castillo et al. \cite{Castillo2011EvaluationOT}.

\textbf{Python frameworks \& energy:} The prevalence of Python in scientific computing, ML, and HPC motivates energy analysis in this context. Some studies have shown that the choice of library (e.g., Pandas vs. Polars \cite{Nahrstedt2024}) and optimization strategies \cite{Holm2020} can have a significant impact on the energy efficiency of Python codes, particularly on GPUs. With the rise of deep learning, performance and energy comparisons between frameworks like TensorFlow and PyTorch on various accelerators became common \cite{Wang2020, Georgiou2022}, highlighting the effect of the framework and the hardware on energy efficiency.

\textbf{Our contribution:} Our work cover each of the previous aspects by  evaluating the impact of hardware-level power management techniques (DVFS, Power Capping, ACPI) on energy efficiency through a CPU/GPU benchmarking with Python-based frameworks (JAX and TensorFlow). Unlike studies focusing purely on framework comparison or hardware modelling, we provide a systematic empirical analysis of how users can modulate energy and performance using system controls at the software-level.

\section{Technical foundations and benchmarking}\label{sec:experimental}
\subsection{Background}

\subsubsection{ACPI/P-State governor scaling mode}
Advanced Configuration and Power Interface (ACPI) is an open framework that Operating Systems can use to discover/configure computer hardware components and to manage/monitor power consumption. Within this framework, the CPU frequency scaling is handled through {\em ACPI/P-State governors}, which dynamically adjust CPU frequency and voltage based on the current status and workload. The most common governors include:
\begin{itemize}
    \item \textbf{Performance}: CPU is kept at highest frequency, thus maximizing performance regardless of power drawn.
    \item \textbf{Powersave}: CPU is kept at its lowest frequency, thus minimizing power consumption regardless of performance.
    \item \textbf{Ondemand}:  CPU frequency is dynamically adjusted based on the system load, with the aim of balancing between power consumption and performance.
    \item \textbf{Conservative}: Similar to Ondemand but with a slower frequency scaling rate, resulting in smoother transitions between different frequencies.
    \item \textbf{Schedutil}: A newer governor that integrates CPU frequency scaling with the Linux kernel's scheduler. CPU frequency is adjusted based on task scheduling and workload, providing more responsive and efficient scaling compared to \textit{Ondemand} and \textit{Conservative}.
    \item \textbf{Userspace}: This governor allows to manually set the CPU frequency, which is useful in case specific frequency settings are needed or wished.
\end{itemize}
\centerline

\subsubsection{Power Limitation or Capping}\
\\
{\em Power limitation}, also known as {\em power capping}, is a technique used to bound power consumption through controlling its maximum. By turning on a power cap, the system enforces a limit on the power drawn by the processor, ensuring that it does not exceed a predefined threshold. This technique is especially useful with data centers, where managing power budgets is crucial for operational efficiency and for preventing overheating. Power capping can be implemented at the hardware level using built-in features like Intel's RAPL or through software-based solutions that monitor and adjust power usage dynamically.
\centerline

\subsubsection{Frequency limitation on CPU and GPU}\
\\
{\em Frequency limitation} involves bounding the clock frequency of the processor (i.e. setting maximum and/or minimum) in order to control power consumption and thermal effects, considering the well-known correlation between  power consumption, frequency and voltage ($P=CV^{2}F$). For CPUs (resp. GPUs), frequency limitation can be achieved through BIOS settings or operating system-level tools such as \texttt{cpufreq} on Linux (resp. NVIDIA’s \texttt{nvidia-smi}). A skillful management of frequency limits that can significantly reduce the overall power consumption as the computing power can thereby be adapted to the effective need of the workloads.

\subsection{Experimental testbed description}\label{sec:method}
For energy measurement, we used EA2P (\textit{Energy-Aware Application Profiler}) \cite{ea2p}, a CPU/GPU energy profiler tool. Considering Python codes allows to leverage popular GPU-accelerated frameworks such as \texttt{JAX} and \texttt{TF} for our study. Each experiment is repeated five times to ensure consistency and to take into account the potential fluctuation coming from the operating system (OS) activity, then the mean value is calculated and considered.\\

\subsubsection{Frameworks}\
\\
\textbf{JAX:} Combines NumPy-like API with automatic differentiation and \textit{Just-In-Time (JIT)} compilation via the XLA (Accelerated Linear Algebra) compiler. XLA performs aggressive optimizations like operator fusion, aiming for high performance on accelerators. JAX benchmarks used specific XLA environment variables \textit{(XLA\_PYTHON\_CLIENT\_PREALLOCATE=false, XLA\_PYTHON\_CLIENT\_MEM\_FRACTION=.10, XLA\_PYTHON\_CLIENT\_ALLOCATOR=platform)}. These specific JAX flags were employed to mitigate Out-Of-Memory (OOM) errors encountered with default settings on the A100 for the large target data sizes, enabling execution near the GPU memory capacity. The minimum possible fraction limit was 10\% as we used. However, we observed compiler warnings (\textit{"XLA HLO rematerialization``)} for certain large JAX workloads (Dist and Stencil), indicating operation near memory capacity limits under our configuration.

\textbf{TensorFlow:} A widely adopted framework supporting both eager execution and graph-based execution via the \textit{@tf.function} decorator, which traces Python code to build optimizable computation graphs. It leverages backend libraries like MKL-DNN, cuDNN, and TensorRT for kernel execution. We note that TensorRT libraries were not installed in the experimental environment. So TensorFlow execution proceeded without TensorRT optimizations, as indicated by runtime warnings ('Could not find TensorRT'). TensorFlow-TensorRT (TF-TRT) is an integration that allows TensorFlow to automatically optimize compatible parts of its computation graph using NVIDIA's TensorRT library. Similarly to JAX we observed OOM error in default settings and decide to use TF\_GPU\_ALLOCATOR = 'cuda\_malloc\_async'. This allocator leverages GPU hardware features to make memory allocation and deallocation much faster (often non-blocking for the CPU) and potentially reduces fragmentation compared to the older default synchronous cudaMalloc or BFC allocators.

JAX Version is 0.4.31, TensorFlow version is 2.12.0 and The NVIDIA driver's CUDA version is 12.2. All benchmarks used single-precision floating-point (FP32) arithmetic. \\

\subsubsection{Benchmarks selection}\
\\
We consider six computing kernels with different structures \textit{(compute-bound, memory-bound, or mixed-bound operations)}. 
\begin{itemize}
\item \textit{GEMM} ({\em compute-bound}): Matrix multiplication (NxN); Size (A100 GPU): N = 41000 for JAX and N = 59000 for TF; Size (CPU): N = 20000; Framework-specific kernels used; Iterations: 1 for TF and 2 for JAX (on GPU). 
\item \textit{Stencil}  ({\em mixed}): 7-point 3D stencil (NxNxN); Size (A100/CPU): N = 1500 for JAX and 1000 for TF; Iterations: 10 for CPU (both TF and JAX) and 50 for GPU (both TF and JAX). 
\item \textit{SpMV}  ({\em memory-bound}): Sparse Matrix-Vector product; GPU Only; A100 JAX: N=38000, density=0.05 (BCOO format); A100 TF: N = 110000, density=0.05 (SparseTensor);  Iterations: 100 for both frameworks.
\item \textit{Triad} ({\em memory-bound}): STREAM Triad variant on vectors of size N; Size (A100/CPU): $N=2\times10^9$; Iterations: 20 for CPU and 100 for A100.
\item \textit{Dist} ({\em memory-bound}): Euclidean distance calculation on vectors of size N. Size (A100/CPU): $N=2\times10^9$ (TF) / $3\times10^9$ (JAX); Iterations: 10 for CPU (both TF and JAX), 100 for TF and 1000 for JAX on A100.
\item \textit{Monte Carlo} ({\em mixed}): Pi estimation using N points; Size (A100/CPU): $N=2\times10^9$; Iterations: JAX = 50, TF = 10 on CPU and JAX = 500, TF = 100. 
\end{itemize}
Benchmark parameters (N, iterations) for the A100 were chosen to maximize GPU utilization and memory occupancy within the 40GB limit for each framework so as to get more realistic power behaviour. We increased the number of iterations for very fast kernels in order to ensure that the measurement duration will exceed the tool's sampling interval. This occasionally led to differing configurations between frameworks for the same benchmark.

\centerline

\subsubsection{Platforms and Configurations}\
\\
Table \ref{tab:sys_charac} displays the hardware characteristics of the two platforms we considered for our study (\cite{epyc_specs}, \cite{xeon_specs}).

\begin{table}[!htbp]
    \centering
    \caption{Platform Characteristics}
    \label{tab:sys_charac}
    \begin{tabular}{|l|c|c|c|} 
    \hline
    \textbf{Name} & {\bf Intel} & {\bf AMD/Nvidia} \\ 
    \hline
    \textbf{CPU model} & Platinum 8358 (x2) & EPYC 7513 (x1) \\ 
    \hline
    \textbf{Clockspeed} & 2.6 GHz & 2.6 GHz \\	
    \hline
    \textbf{Turbo Speed} & Up to 3.4 GHz & Up to 3.65 GHz \\  
    \hline
    \textbf{Cores/Threads} & 32/64 (x2) & 32/64 \\  
    \hline	
    \textbf{L1 iCache} & 1,024KB 8-way & 1,024KB 8-way \\  
    \hline
    \textbf{L1 dCache} & 1,536KB 12-way & 1,024KB 8-way \\  
    \hline
    \textbf{L2 Cache} & 40MB 20-way & 16MB 8-way \\  
    \hline
    \textbf{L3 Cache} & 48MB 12-way & 128MB 16-way \\  
    \hline
    \textbf{DRAM Memory} & 512GB DDR4-3200 & 512GB DDR4-3200\\
    \hline
    \textbf{CPU TDP} & 250W (x2) & 200W \\  
    \hline
    \textbf{GPU Model} & / & Nvidia A100 SXM4\\  
    \hline
    \textbf{GPU TDP} & / & 400W \\  
    \hline
    \textbf{GPU Memory} & / & 40GB HBM2 \\  
    \hline
    \textbf{Data precision} & Single & Single \\  
    \hline
    \textbf{SIMD extensions} & SSE, AVX, AVX512 & SSE, AVX \\  
    \hline
    \textbf{Operating System} & Debian 5.10.209 & Debian 5.10.209 \\
    \hline
    \end{tabular}
\end{table}


For each platform, a specific set of configurations is considered for the management of power consumption and performance. These configurations are detailed in Table \ref{tab:power_settings}.

\begin{table}[!htbp]
    \centering
    \footnotesize
    \setlength{\tabcolsep}{1.0pt}
    \caption{Combined DVFS, ACPI P-States, and Power Cap Settings}
    \label{tab:power_settings}
    \begin{tabular}{|c|c|c|c|}
    \hline
    \textbf{Setting num} & \textbf{Ice Lake SP} & \textbf{Zen 3} & \textbf{NVIDIA A100} \\ \hline
     & \multicolumn{3}{c|}{\textbf{DVFS (Frequency in MHz)}} \\ \hline
    1 & 3400 & 3600 & 1215 (Mem), 1410 (GPU) \\ \hline
    2 & 3000 & 3300 & 1215 (Mem), 1215 (GPU) \\ \hline
    3 & 2700 & 3000 & 1215 (Mem), 1005 (GPU) \\ \hline
    4 & 2400 & 2700 & 1215 (Mem), 810 (GPU) \\ \hline
    5 & 2100 & 2400 & 1215 (Mem), 600 (GPU) \\ \hline
    6 & 1800 & 2100 & 1215 (Mem), 405 (GPU) \\ \hline
    7 & 1500 & 1800 & 1215 (Mem), 210 (GPU) \\ \hline
    8 & 1200 & 1500 & \\ \hline
    9 & 800 & & \\ \hline
    & \multicolumn{3}{c|}{\textbf{ACPI P-States (Governor)}} \\ \hline
    1 & Performance & Powersave & \\ \hline
    2 & Powersave & Performance & \\ \hline
    3 &  & Ondemand  & \\ \hline
    4 &  & Schedutil & \\ \hline
    5 &  & Userspace & \\ \hline
    6 &  & Conservative & \\ \hline
    & \multicolumn{3}{c|}{\textbf{Power Cap (Power Limit)}} \\ \hline
    1 & 100 W &  & 100 W \\ \hline
    2 & 150 W &  & 150 W \\ \hline
    3 & 200 W &  & 200 W \\ \hline
    4 & 250 W &  & 250 W \\ \hline
    5 &  &  & 300 W \\ \hline
    6 &  &  & 350 W \\ \hline
    7 &  &  & 400 W \\ \hline
    \end{tabular}
\end{table}

Our settings related to frequency/power limitation range from the minimum to maximum allowed by the system (for example: [800MHz; 3400MHz] for our Intel CPU frequency and  [100W; 400W] for our Nvidia GPU power).
\centerline


\subsubsection{Metrics and measurement}\
\\
To consistently evaluate time and energy performance under different configurations by collecting detailed time and power data using our custom \textit{ea2p} tool \cite{ea2p}. For each scenario, our benchmarking considers several runs (so-called {\em iteration}) followed by the calculation of the mean value of collected measurements. The key metrics are:  \textit{total execution time},  \textit{component Power (CPU Pkg, GPU and DRAM)}, \textit{total energy consumed}, and \textit{Energy-Delay Product (EDP)}. While data related to {\em  time} and {\em energy} are measurements, values for {\em power} and {\em EDP} are calculated with the corresponding energy and time measurements (i.e $P = E/T$ and $EDP=E\times T$).



\section{Results and analysis}\label{sec:analysis}
We first establish the {\em baseline} performance and energy characteristics under standard  configurations. Subsequently, we analyze the impact of applying {\em frequency limitation} (DVFS), {\em OS-level CPU Governors} (ACPI), and {\em Power Capping} (PowerCap) techniques on {\em execution time}, {\em component power draw}, {\em system energy consumption}, and the {\em Energy-Delay Product} (EDP). We investigate trends across benchmarks and platforms, and analyze the behavior of the JAX and TensorFlow frameworks.
\centerline

\subsection{Baseline characteristics}
To establish a reference point for evaluating energy management techniques, we first characterized the performance and energy consumption under default high-performance configurations (i.e; frequency at the highest) for each platform. For our CPU platforms, this corresponds to the ACPI 'performance' governor (Intel: Setting 1, AMD: Setting 2). For Nvidia A100 GPU, the baseline uses the highest manually configured DVFS setting (Setting 1: MEM 1215 MHz, GRAPH 1410 MHz). Table \ref{tab:baseline} summarizes three key metrics (mean ± standard deviation over 5 repetitions): {\em time, power} and {\em energy}  from our benchmarking under baseline settings.\\

\subsubsection{Execution time}\ 
\\
As shown in Table \ref{tab:baseline}, baseline execution times vary significantly based on the platform, benchmark, and framework. The Nvidia A100 platform generally achieved the lowest execution times, often sub-second for workloads like Triad, Dist, and Monte Carlo, reflecting its high parallelism and memory bandwidth. The CPU platforms exhibited longer execution times, particularly for compute-intensive tasks like GEMM (e.g., $>8$ seconds on AMD, $2.5-4.6$ seconds on Intel). Memory-bound workloads like Triad were faster on CPU ($\approx 0.3-0.7$ seconds). Comparing frameworks at baseline requires caution, as workload configurations (data sizes, iterations) sometimes differed (caveat applies to respective rows in Table \ref{tab:baseline}).

\begin{table}[!htbp]
    \centering
    \footnotesize
    \setlength{\tabcolsep}{2.5pt}
    \caption{Baselines results for all platforms. 
    }
    \label{tab:baseline}
    \begin{tabular}{|l|l|l|l|rcl|l|}
        \toprule
        Platform & App & Lib & Time (s) & \multicolumn{3}{c|}{Energy (J)} & Power (W) \\
        \midrule
        intel & dist* & jax & 0.873 ± 0.028 & 273.3 &±& 5.4 & 313.7 ± 3.7 \\
        intel & dist* & tf & 0.555 ± 0.001 & 301.1 &±& 0.8 & 543.5 ± 0.9 \\
        intel & gemm & jax & 4.625 ± 0.222 & 2542.5 &±& 136.3 & 551.1 ± 13.3 \\
        intel & gemm & tf & 2.517 ± 0.001 & 1315.4 &±& 8.6 & 524.8 ± 3.3 \\
        intel & m\_c* & jax & 0.538 ± 0.009 & 85.5 &±& 0.7 & 318.4 ± 3.8 \\
        intel & m\_c* & tf & 1.204 ± 0.001 & 666.5 &±& 1.3 & 553.9 ± 1.3 \\
        intel & stencil* & jax & 1.580 ± 0.027 & 491.1 &±& 7.4 & 310.4 ± 1.2 \\
        intel & stencil* & tf & 1.290 ± 0.027 & 552.2 &±& 5.2 & 429.3 ± 6.4 \\
        intel & triad & jax & 0.431 ± 0.011 & 136.8 &±& 1.8 & 317.7 ± 5.0 \\
        intel & triad & tf & 0.528 ± 0.000 & 301.1 &±& 1.2 & 570.3 ± 1.9 \\
        amd & dist* & jax & 0.669 ± 0.026 & 104.6 &±& 2.7 & 156.7 ± 2.6 \\
        amd & dist* & tf & 1.323 ± 0.024 & 211.5 &±& 3.5 & 160.0 ± 2.7 \\
        amd & gemm & jax & 8.007 ± 0.141 & 1420.4 &±& 28.9 & 177.7 ± 3.2 \\
        amd & gemm & tf & 8.058 ± 0.070 & 1380.3 &±& 27.8 & 171.5 ± 4.3 \\
        amd & m\_c* & jax & 0.400 ± 0.011 & 62.2 &±& 1.4 & 155.5 ± 2.2 \\
        amd & m\_c* & tf & 2.349 ± 0.013 & 460.7 &±& 2.8 & 196.2 ± 0.5 \\
        amd & stencil* & jax & 0.712 ± 0.029 & 118.6 &±& 3.4 & 166.6 ± 3.6 \\
        amd & stencil* & tf & 2.350 ± 0.035 & 373.0 &±& 2.4 & 159.0 ± 2.5 \\
        amd & triad & jax & 0.324 ± 0.016 & 49.4 &±& 1.9 & 152.5 ± 2.6 \\
        amd & triad & tf & 0.681 ± 0.027 & 112.3 &±& 3.8 & 165.2 ± 3.6 \\
        nvidia & dist*/** & jax & 0.016 ± 0.000 & 5.6 &±& 0.0 & 342.7 ± 3.4 \\
        nvidia & dist*/** & tf & 0.063 ± 0.000 & 19.8 &±& 0.3 & 316.7 ± 6.4 \\
        nvidia & gemm*/** & jax & 1.442 ± 0.007 & 716.4 &±& 51.2 & 498.6 ± 36.6 \\
        nvidia & gemm*/** & tf & 3.961 ± 0.006 & 1953.6 &±& 130.8 & 494.5 ± 32.6 \\
        nvidia & m\_c*/** & jax & 0.011 ± 0.000 & 3.9 &±& 0.1 & 343.3 ± 7.2 \\
        nvidia & m\_c*/** & tf & 0.079 ± 0.000 & 28.0 &±& 1.2 & 356.7 ± 14.6 \\
        nvidia & spmv*/** & jax & 0.045 ± 0.000 & 8.8 &±& 0.1 & 196.8 ± 2.3 \\
        nvidia & spmv*/** & tf & 0.034 ± 0.000 & 9.4 &±& 0.1 & 276.4 ± 2.1 \\
        nvidia & stencil*/** & jax & 0.405 ± 0.006 & 48.8 &±& 2.5 & 120.5 ± 6.3 \\
        nvidia & stencil*/** & tf & 0.260 ± 0.000 & 81.9 &±& 0.7 & 320.4 ± 2.8 \\
        nvidia & triad & jax & 0.227 ± 0.000 & 25.4 &±& 1.2 & 111.7 ± 5.3 \\
        nvidia & triad & tf & 0.035 ± 0.000 & 11.1 &±& 0.3 & 320.6 ± 6.9 \\
        \bottomrule
    \end{tabular}
\end{table}
\centerline

\subsubsection{Power and energy}\ 
\\
System power draw (CPU+DRAM for Intel and AMD; CPU+GPU+DRAM for Nvidia) at baseline also shows noticeable variations. The dual-socket Intel Xeon platform consistently hits the highest power, frequently exceeding 500 W for most demanding workloads (especially TF), while still reflecting the indicated TDP of 250 W per socket. Nvidia A100 shows high peak power potential (up to $\approx 500W$ for TF GEMM) with also a noticeable variation depending on workload fraction offloaded to the GPU. The single-socket AMD EPYC platform exhibits the lowest baseline power draw, typically operating below 200W. Consequently, energy seems to align with the aforementioned {\em power levels} and the {\em running times}; faster runs with A100 more often consume less energy despite a high running power, while long lasting CPU runs show significant energy values.

Our baseline results illustrates the important amount of energy associated with high-performance processing, thus motivating our investigation of energy management techniques. Observed performance/energy gaps between different platforms/frameworks provide useful data for experimentally figure out the impact of DVFS, ACPI, and Power Capping.

\subsection{EDP analysis}
\begin{figure*}[htbp]
    \centering
    \includegraphics[scale=0.27]{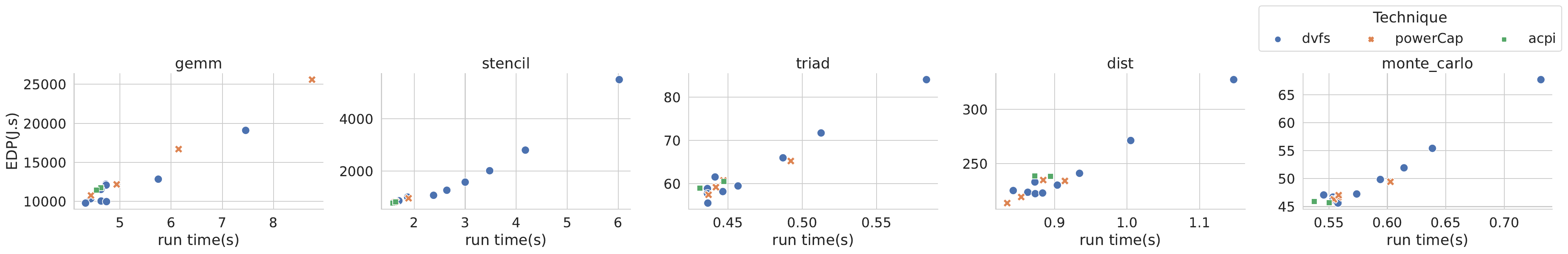}
    \caption{EDP vs Time on Intel for JAX}
    \label{fig:intel_jax_edp}
\end{figure*}

\begin{figure*}[htbp]
    \centering
    \includegraphics[scale=0.27]{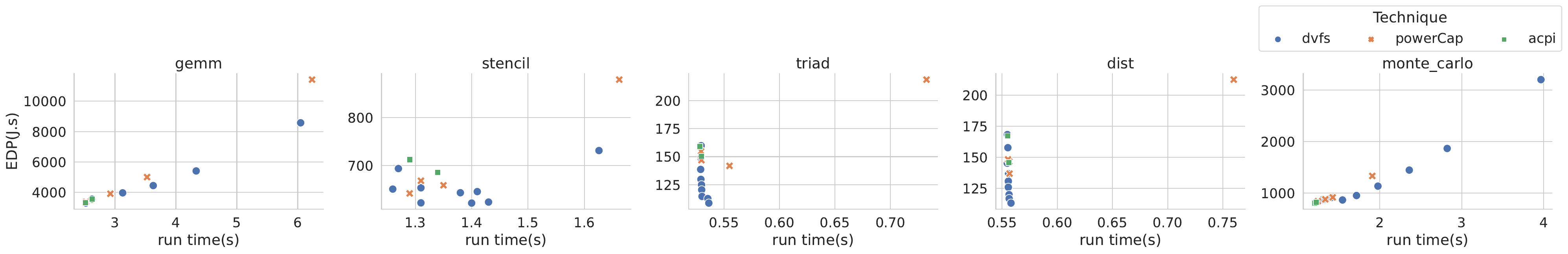}
    \caption{EDP vs Time on Intel for TF}
    \label{fig:intel_tf_edp}
\end{figure*}

\begin{figure*}[htbp]
    \centering
    \includegraphics[scale=0.27]{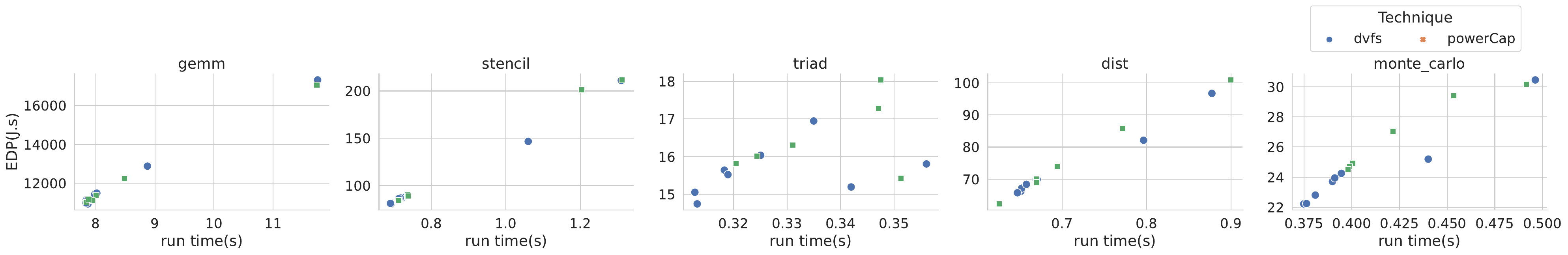}
    \caption{EDP vs Time on AMD for JAX}
    \label{fig:amd_jax_edp}
\end{figure*}

\begin{figure*}[htbp]
    \centering
    \includegraphics[scale=0.27]{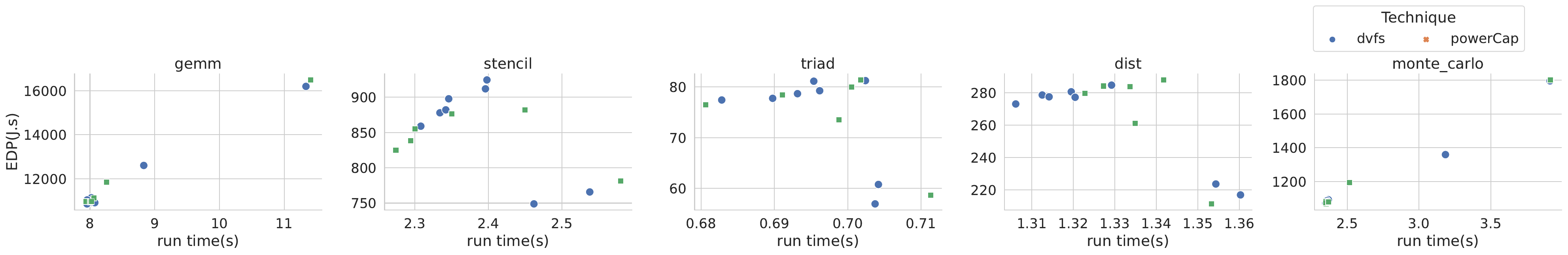}
    \caption{EDP vs Time on AMD for TF}
    \label{fig:amd_tf_edp}
\end{figure*}

\begin{figure*}[htbp]
    \centering
    \includegraphics[scale=0.23]{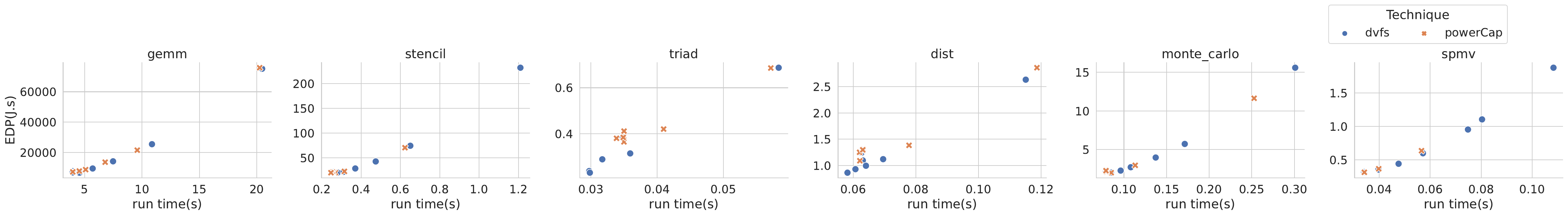}
    \caption{EDP vs Time on Nvidia for TF}
    \label{fig:nvidia_tf_edp}
\end{figure*}

The plots of the EDP vs time (Figures \ref{fig:intel_jax_edp}, \ref{fig:intel_tf_edp}, \ref{fig:amd_jax_edp}, \ref{fig:amd_tf_edp},   \ref{fig:nvidia_jax_edp} and \ref{fig:nvidia_tf_edp}) quantify the efficiency trade-offs achievable with different energy management techniques. Tables  \ref{tab:best_EDP_intel_JAX}, \ref{tab:best_EDP_intel_TF}, \ref{tab:best_EDP_amd_JAX}, \ref{tab:best_EDP_amd_TF}, \ref{tab:best_EDP_nvidia_JAX} and \ref{tab:best_EDP_nvidia_TF} summarize the optimal EDP observed for each instance of (\{)\textit{Platform, Framework, Benchmark, ManagementMethod}), presenting (in percentage) EDP reduction and the corresponding running time gap relative to the baseline (provided in \ref{tab:baseline}). Direct comparison between the aforementioned values should consider the content of both Table \ref{tab:baseline} and Section \ref{sec:method}.\\

\subsubsection{General observations}\ 
\\
The plots confirm that some configurations of out the baseline ones yield a significant reduction of the EDP, each management technique having a specific impact. Smallest EDP  often occur with settings that involve reduced frequency or limited power. Some cases show a performance penalty as the price for a greater energy savings. However, applying extreme energy-saving settings does not always yield the best EDP, as excessive increases in execution time could outweigh the corresponding energy reduction.\\

\subsubsection{Platform specific observations}\
\\
\textbf{Intel:} Power capping frequently yields substantial EDP reductions (e.g., 10-18\% for TF Dist/Triad, 8-10\% for JAX GEMM/Dist) with slightly lower running times, thus indicating its potential for EDP efficiency without a significant performance loss. DVFS also lowers EDP (up to $\approx32\%$ for TF Triad/Dist), but the optimal EDP often comes  with important performance degradation ($>0\%$ time increase). The ACPI \textit{'powersave'} governor generally yields slightly higher EDP than the \textit{'performance'} baseline.

\textbf{AMD:} For many of our workloads (Stencil, Triad, Dist, Monte Carlo), DVFS consistently yields lowest EDP values (4-11\% for JAX and 1-25\% for TF), often at the expense of the running time (especially for TF). About ACPI governors, which provide distinct operating points, \textit{'powersave'} significantly reduces EDP compared to \textit{'performance'} but with a substantial increase of the running time, while DVFS allows finer tuning to reach the optimal EDP setting.

\textbf{Nvidia:} Power Capping is highly effective for EDP reduction on A100. For JAX, it yields significant EDP reductions (34-42\%) across several benchmarks (Dist, MC, SpMV, Stencil, Triad) with relatively small time increases (0-2\%). For TF, Power Capping also reduces EDP (up to $\approx32\%$ for Dist/MC), often more effectively than DVFS. DVFS reduces EDP (notably with GEMM and TF Triad) but seems limited by JAX's inability to run at the lowest frequencies (Settings 6, 7), sometimes incurring larger time penalties than PowerCap for similar EDP gains.

\begin{figure*}[htbp]
    \centering
    \includegraphics[scale=0.23]{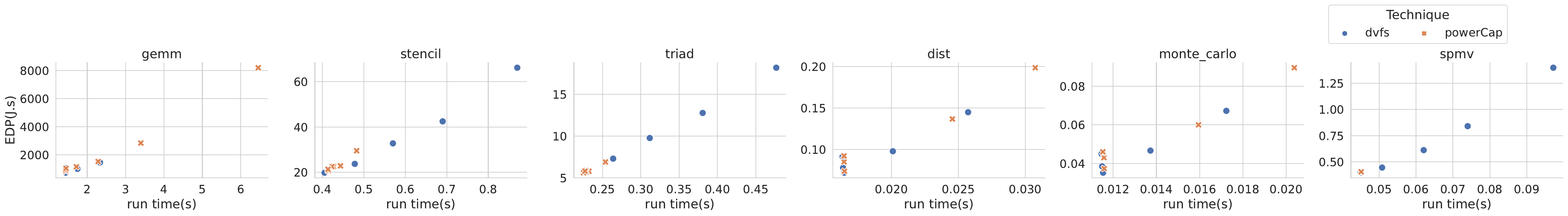}
    \caption{EDP vs Time on Nvidia for JAX}
    \label{fig:nvidia_jax_edp}
\end{figure*}

In summary, our power analysis at the component level reveals that DVFS and Power Capping primarily modulate the overall CPU/GPU power consumption. JAX and TF show  different energy efficiencies with similar power draw values and different running times.\\

\subsubsection{Workload sensitivity}\ 
\\
The optimal strategy depends on the nature of the workload. GEMM, which is compute-bound GEMM, gets its minimum EDP with less aggressive settings (higher frequencies/power caps) compared to memory-bound workloads like Triad or Dist for which lower DVFS frequencies often give optimal EDP. On A100, the effectiveness of Power capping even for compute-heavy tasks like GEMM (16-34\% EDP reduction via PowerCap on Nvidia) illustrates its viability across different workload types, provided not too low caps.

\begin{table*}[htbp]
    \centering
    \footnotesize
    \setlength{\tabcolsep}{1.0pt}
    \caption{Best EDP improvement with associated execution time on Intel with JAX}
    \label{tab:best_EDP_intel_JAX}
    \begin{tabular}{|l||r@{\hskip 0.1in}r@{\hskip 0.1in}c|r@{\hskip 0.1in}r@{\hskip 0.1in}c|r@{\hskip 0.1in}r@{\hskip 0.1in}l|}
        \hline
        Bench &  \multicolumn{3}{c|}{\texttt{DVFS}} &  \multicolumn{3}{c|}{\texttt{POWERCAP}} &  \multicolumn{3}{c|}{\texttt{ACPI}}\\
         & EDP&Time&Freq (Ghz) & EDP&Time&Power (W) & EDP&Time&Mode \\
        \hline
        \hline
   dist & +6.9\% & 0.0\% & 2.4 & +10.5\% & -4.4\% & 250 & +0.1\% &+2.5\% & powersave \\
        gemm & +16.7\% & -6.4\% & 2.1 & +8.4\% & -4.1\% & 250  & +2.5\% & -1.8\% & powersave \\
        m\_c & +0.7\% & +3.7\% & 2.4 & -0.8\% & +3.1\% & 200 & +0.5\% & +2.4\% & powersave \\
        stencil & -2.9\% & +1.3\% & 3.4  & -4.3\% & +2.3\% & 250 & $\pm$0.0\% & 0.0\% & performance \\
        triad & +5.9\% & +1.2\% & 2.1 & +2.6\% & +1.3\% & 150   & $\pm$0.0\% & 0.0\% & performance\\
        \hline
    \end{tabular}
\end{table*}

\begin{table*}[htbp]
    \centering
    \footnotesize
    \setlength{\tabcolsep}{1.0pt}
    \caption{Best EDP improvement with associated execution time on Intel with TF}
    \label{tab:best_EDP_intel_TF}
    \begin{tabular}{|l||r@{\hskip 0.1in}r@{\hskip 0.1in}c|r@{\hskip 0.1in}r@{\hskip 0.1in}c|r@{\hskip 0.1in}r@{\hskip 0.1in}l|}
        \hline
        Bench &  \multicolumn{3}{c|}{\texttt{DVFS}} &  \multicolumn{3}{c|}{\texttt{POWERCAP}} &  \multicolumn{3}{c|}{\texttt{ACPI}}\\
         & EDP&Time&Freq (Ghz) & EDP&Time&Power (W) & EDP&Time&Mode \\
        \hline
        \hline
        dist & +32.3\%  & +0.5\% & 0.8 & +18.1\%  & +0.3\% & 150& +12.8\%  & +0.2\% & powersave\\
        gemm & +0.8\%  & 0.0\% & 2.7 & -1.5\%  & +0.3\% & 250& 0.0\%  & 0.0\% & performance \\
        m\_c & -1.5\%  & +0.9\% & 3.4 & -4.7\%  & +3.4\% & 250& 0.0\%  & 0.0\% & performance \\
        stencil & +12.7\%  & +8.5\% & 1.5 & +9.8\%  & 0.0\% & 250& +3.7\%  & +3.9\% & powersave \\
        triad & +31.8\%  & +1.5\% & 0.8 & +10.9\%  & +5.0\% & 150& +5.4\%  & +0.3\% & powersave \\
        \hline
    \end{tabular}
\end{table*}

\begin{table}[htbp]
    \centering
    \footnotesize
    \setlength{\tabcolsep}{1.0pt}
    \caption{Best EDP improvement with associated execution time on AMD with JAX}
    \label{tab:best_EDP_amd_JAX}
    \begin{tabular}{|l||r@{\hskip 0.1in}r@{\hskip 0.1in}c|r@{\hskip 0.1in}r@{\hskip 0.1in}l|}
        \hline
        Bench &  \multicolumn{3}{c|}{\texttt{DVFS}} & \multicolumn{3}{c|}{\texttt{ACPI}}\\
         & EDP&Time&Freq (Ghz) & EDP&Time&Mode \\
        \hline
        \hline
        dist & +6.1\% & -3.3\% & 2.4 & +11.1\% & -6.6\% & ondemand\\
        gemm & +3.9\% & -1.8\% & 3.3 & +3.1\% & -2.2\% & schedutil\\
        m\_c & +10.8\% & -6.4\% & 2.7 & +1.7\% & -0.6\% & schedutil\\
        stencil & +4.5\% & -2.8\% & 3.6 & 0.0\% & 0.0\% & performance\\
        triad  & +7.9\% & -3.4\% & 2.7 & +3.7\% & +8.3\% & powersave\\
        \hline
    \end{tabular}
\end{table}

\begin{table}[htbp]
    \centering
    \footnotesize
    \setlength{\tabcolsep}{1.0pt}
    \caption{Best EDP improvement with associated execution time on AMD with TF}
    \label{tab:best_EDP_amd_TF}
    \begin{tabular}{|l||r@{\hskip 0.1in}r@{\hskip 0.1in}c|r@{\hskip 0.1in}r@{\hskip 0.1in}l|}
        \hline
        Bench &  \multicolumn{3}{c|}{\texttt{DVFS}} & \multicolumn{3}{c|}{\texttt{ACPI}}\\
         & EDP&Time&Freq (Ghz) & EDP&Time&Mode \\
        \hline
        \hline
        dist & +22.5\% & +2.8\% & 1.5 & +24.5\% & +2.3\% & powersave\\
        gemm & +2.4\% & -1.3\% & 2.1 & +1.6\% & -0.6\% & schedutil\\
        m\_c & +1.0\% & 0.0\% & 2.1  & +1.4\% & +0.2\% & ondemand\\
        stencil & +14.6\% & +4.8\% & 1.8 & +10.9\% & +9.8\% & powersave\\
        triad & +25.5\% & +3.4\% & 1.5 & +23.3\% & +4.5\% & powersave\\
        \hline
    \end{tabular}
\end{table}

\begin{table}[htbp]
    \centering
    \footnotesize
    \setlength{\tabcolsep}{1.0pt}
    \caption{Best EDP improvement with associated execution time on Nvidia with JAX}
    \label{tab:best_EDP_nvidia_JAX}
    \begin{tabular}{|l||r@{\hskip 0.1in}r@{\hskip 0.1in}c|r@{\hskip 0.1in}r@{\hskip 0.1in}l|}
        \hline
        Bench &  \multicolumn{3}{c|}{\texttt{DVFS}} & \multicolumn{3}{c|}{\texttt{POWERCAP}}\\
         & EDP&Time&Freq (Ghz) & EDP&Time&Power (W) \\
        \hline
        \hline
        dist & +21.4\% & +0.8\% & 1.005& +37.4\% & +0.9\% & 200 \\
        gemm & +30.6\% & +0.1\% & 1.005 & +34.2\% & 0.0\% & 300 \\
        m\_c & +21.6\% & +0.6\% & 1.005 & +37.4\% & +1.1\% & 200 \\
        spmv & 0.0\% & 0.0\% & 1.410 & +34.1\% & +0.4\% & 100\\
        stencil  & 0.0\% & 0.0\% & 1.410 & +37.3\% & +2.3\% & 300 \\
        triad & 0.0\% & 0.0\% & 1.410 & +41.8\% & -0.9\% & 300 \\
        \hline
    \end{tabular}
\end{table}

\begin{table}[htbp]
    \centering
    \footnotesize
    \setlength{\tabcolsep}{1.0pt}
    \caption{Best EDP improvement with associated execution time on Nvidia for TF}
    \label{tab:best_EDP_nvidia_TF}
    \begin{tabular}{|l||r@{\hskip 0.1in}r@{\hskip 0.1in}c|r@{\hskip 0.1in}r@{\hskip 0.1in}l|}
        \hline
        Bench &  \multicolumn{3}{c|}{\texttt{DVFS}} & \multicolumn{3}{c|}{\texttt{POWERCAP}}\\
         & EDP&Time&Freq (Ghz) & EDP&Time&Power (W) \\
        \hline
        \hline
        dist & +30.8\% & -7.2\% & 1.005 & +32.5\% & -0.9\% & 200\\
        gemm & +12.8\% & +15.0\% & 1.005 & +16.3\% & +14.5\% & 300\\
        m\_c & +8.6\% & +7.6\% & 1.215 & +32.4\% & +8.1\% & 200\\
        spmv & 0.0\% & 0.0\% & 1.410 & +24.2\% & +0.3\% & 350\\
        stencil & +6.5\% & +7.7\% & 1.215 & +29.9\% & +3.2\% & 200\\
        triad & +40.5\% & -14.2\% & 0.810 & +23.9\% & +0.6\% & 200\\
        \hline
    \end{tabular}
\end{table}
\centerline

\subsubsection{Framework efficiency trends}\ 
\\
Considering Triad on A100, Power capping yields better EDP reduction for JAX (42\%) than for TF (24\%), with a significant improvement over DVFS for both. For other benchmarks,  EDP savings should be interpreted should consider the workload differences. JAX's inability to handle the lowest A100 DVFS settings had limited the achievable EDP range with that specific method.

\subsection{Specific impacts of power management techniques}
Figures \ref{fig:intel_power}, \ref{fig:amd_power} and \ref{fig:nvidia_power} display the (average) {\em power draw} of the primary component (CPU Package 'Pkg' for Intel/AMD, 'GPU' for Nvidia) and a secondary component ('DRAM' for CPUs, 'Host' CPU for Nvidia) with our considered management settings. Tables  \ref{tab:baseline_power_EDP_amd}, \ref{tab:baseline_power_EDP_intel} and \ref{tab:baseline_power_EDP_nvidia} provide the specific power values (mean ± std) which respect to the baseline and the minimum EDP.\\

\subsubsection{Impact of the DVFS}\
\\
Looking at the top-half part of Figures \ref{fig:intel_power} and \ref{fig:nvidia_power} and bottom-half part of Figure \ref{fig:amd_power}, we can see that applying DVFS consistently and significantly reduces the power consumption of the main compute component (Pkg on CPUs, GPU on A100). The reduction is particularly significant for compute-intensive workloads like GEMM, which utilize the compute units more intensively at baseline. For memory-bound workloads (e.g., Triad, Dist), the baseline compute power is lower and the reduction with frequency scaling is less pronounced, thus suggesting that these workloads are less constrained by CPU frequency. Comparing baseline vs. Min EDP power (Table \ref{tab:baseline_power_EDP_intel}, \ref{tab:baseline_power_EDP_amd} and \ref{tab:baseline_power_EDP_nvidia}) confirms substantial Pkg/GPU power drops with DVFS, where the EDP is minimal at lower settings.

As can be observed from Tables \ref{tab:baseline_power_EDP_amd}, \ref{tab:baseline_power_EDP_intel}  and \ref{tab:baseline_power_EDP_nvidia}, DRAM power on both Intel and AMD platforms remains low ($\approx 20-80W$) and shows negligible change across DVFS settings. Similarly, the Host CPU power on Nvidia A100 remains consistently low ($<75W$) across DVFS settings. This clearly indicates that CPU/GPU core frequency scaling through  DVFS has an impact on the power draw of the targeted compute unit, with minimal effect on the power of the DRAM or the host CPU (for the GPU).

\begin{table*}[htbp]
    \centering
    \footnotesize
    \setlength{\tabcolsep}{1.0pt}
    \caption{CPU/RAM Baseline power and Min EDP on AMD}
    \label{tab:baseline_power_EDP_amd}
    \begin{tabular}{|ll@{\hskip 0.05in}|@{\hskip 0.05in}r@{\hskip 0.1in}r@{\hskip 0.05in}|@{\hskip 0.05in}r@{\hskip 0.1in}r@{\hskip 0.1in}r@{\hskip 0.1in}r@{\hskip 0.05in}|}
    \hline
         &  & \multicolumn{2}{c}{\texttt{Baseline (W)}}& \multicolumn{4}{c|}{\texttt{MinEDP (W))}} \\
         
     &  & CPU & DRAM  & ACPI CPU & ACPI DRAM& DVFS CPU& DVFS DRAM\\
    Bench & Lib &  &  &  &  &  &  \\
    \hline
    \hline
    \multirow[t]{2}{*}{dist} & jax & 124.5 ± 2.5 & 32.2 ± 0.2 & 127.0 ± 1.8 & 32.5 ± 0.1 & 125.0 ± 1.5 & 32.4 ± 0.3 \\
     & tf & 137.9 ± 2.7 & 22.1 ± 0.4 & 92.5 ± 1.9 & 23.0 ± 1.0 & 93.9 ± 0.7 & 23.5 ± 0.9 \\
    \cline{1-8}
    \multirow[t]{2}{*}{gemm} & jax & 161.5 ± 3.2 & 16.2 ± 0.3 & 163.7 ± 1.1 & 16.1 ± 0.1 & 160.9 ± 5.4 & 16.0 ± 0.2 \\
     & tf & 155.9 ± 4.2 & 15.7 ± 0.1 & 155.3 ± 4.3 & 15.8 ± 0.2 & 156.2 ± 3.1 & 15.6 ± 0.2 \\
    \cline{1-8}
    \multirow[t]{2}{*}{m\_c} & jax & 133.3 ± 1.5 & 22.1 ± 1.0 & 132.8 ± 1.8 & 22.0 ± 0.8 & 136.1 ± 0.9 & 22.2 ± 1.1 \\
     & tf & 173.9 ± 0.6 & 22.3 ± 0.2 & 171.2 ± 2.0 & 21.7 ± 0.4 & 172.2 ± 1.1 & 21.9 ± 0.6 \\
    \cline{1-8}
    \multirow[t]{2}{*}{stencil} & jax & 138.0 ± 3.5 & 28.6 ± 0.5 & 138.0 ± 3.5 & 28.6 ± 0.5 & 139.9 ± 1.9 & 28.7 ± 0.4 \\
     & tf & 140.6 ± 2.5 & 18.3 ± 0.2 & 99.1 ± 1.0 & 18.4 ± 0.9 & 105.5 ± 0.6 & 18.2 ± 0.5 \\
    \cline{1-8}
    \multirow[t]{2}{*}{triad} & jax & 123.8 ± 2.2 & 28.7 ± 0.5 & 96.7 ± 1.7 & 28.7 ± 0.1 & 121.7 ± 3.1 & 29.0 ± 0.1 \\
     & tf & 142.0 ± 3.4 & 23.2 ± 0.6 & 92.9 ± 2.2 & 23.1 ± 0.3 & 91.8 ± 3.1 & 23.3 ± 0.6 \\
    \hline
    \end{tabular}
\end{table*}

\begin{table*}[htbp]
    \centering
    \footnotesize
    \setlength{\tabcolsep}{1.0pt}
    \caption{CPU/RAM Baseline power and Min EDP on Intel. P\_C = POWERCAP}
    \label{tab:baseline_power_EDP_intel}
    \begin{tabular}{|ll@{\hskip 0.05in}|@{\hskip 0.05in}r@{\hskip 0.1in}r@{\hskip 0.05in}|@{\hskip 0.05in}r@{\hskip 0.1in}r@{\hskip 0.1in}r@{\hskip 0.1in}r@{\hskip 0.05in}r@{\hskip 0.1in}r@{\hskip 0.05in}|}
    \hline
         &  & \multicolumn{2}{c}{\texttt{Baseline (W)}}& \multicolumn{6}{c|}{\texttt{MinEDP (W))}} \\
         
     &  & CPU & DRAM  & ACPI CPU & ACPI DRAM& DVFS CPU& DVFS DRAM& P\_C CPU& P\_C DRAM\\
    Bench & Lib &  &  &  &  &  &  &  & \\
    \hline
    \hline
    \multirow[t]{2}{*}{dist} & jax & 259.2 ± 2.7 & 54.6 ± 1.0 & 245.7 ± 4.2 & 52.9 ± 1.8 & 238.7 ± 3.2 & 53.3 ± 0.9 & 252.5 ± 4.6 & 54.9 ± 1.0 \\
     & tf & 457.8 ± 0.7 & 85.7 ± 0.2 & 387.7 ± 9.0 & 85.2 ± 0.2 & 280.7 ± 1.6 & 84.5 ± 0.4 & 355.6 ± 6.2 & 87.3 ± 0.5 \\
    \cline{1-10}
    \multirow[t]{2}{*}{gemm} & jax & 482.4 ± 12.9 & 68.7 ± 1.3 & 489.2 ± 6.7 & 69.2 ± 0.7 & 456.8 ± 17.8 & 68.9 ± 2.4 & 481.5 ± 18.5 & 69.1 ± 2.7 \\
     & tf & 472.6 ± 3.1 & 52.3 ± 0.6 & 472.6 ± 3.1 & 52.3 ± 0.6 & 468.5 ± 4.1 & 52.2 ± 1.2 & 477.7 ± 10.6 & 53.4 ± 1.7 \\
    \cline{1-10}
    \multirow[t]{2}{*}{m\_c} & jax & 262.0 ± 2.7 & 56.3 ± 1.2 & 247.2 ± 9.0 & 55.2 ± 1.6 & 239.2 ± 2.0 & 54.6 ± 0.6 & 246.6 ± 3.6 & 55.5 ± 1.0 \\
     & tf & 493.3 ± 1.2 & 60.6 ± 0.1 & 493.3 ± 1.2 & 60.6 ± 0.1 & 492.0 ± 5.6 & 60.4 ± 0.6 & 482.9 ± 10.4 & 59.5 ± 1.2 \\
    \cline{1-10}
    \multirow[t]{2}{*}{stencil} & jax & 265.7 ± 0.9 & 44.7 ± 0.2 & 265.7 ± 0.9 & 44.7 ± 0.2 & 266.6 ± 1.7 & 44.9 ± 0.3 & 265.8 ± 1.2 & 44.6 ± 0.1 \\
     & tf & 364.0 ± 5.8 & 65.3 ± 0.6 & 319.5 ± 5.1 & 63.3 ± 2.0 & 257.9 ± 7.2 & 60.3 ± 2.5 & 322.2 ± 9.2 & 64.9 ± 2.9 \\
    \cline{1-10}
    \multirow[t]{2}{*}{triad} & jax & 261.1 ± 3.7 & 56.6 ± 1.4 & 261.1 ± 3.7 & 56.6 ± 1.4 & 235.8 ± 3.0 & 56.3 ± 1.0 & 246.3 ± 5.3 & 55.3 ± 1.5 \\
     & tf & 479.9 ± 1.7 & 90.4 ± 0.3 & 446.5 ± 15.7 & 90.4 ± 0.4 & 289.0 ± 2.8 & 89.4 ± 1.3 & 372.0 ± 13.2 & 89.0 ± 0.4 \\
    \cline{1-10}
\end{tabular}
\end{table*}

\begin{table*}[htbp]
    \centering
    \footnotesize
    \setlength{\tabcolsep}{2.0pt}
    \caption{GPU/Host Baseline power and Min EDP on Nvidia. P\_C = POWERCAP}
    \label{tab:baseline_power_EDP_nvidia}
    \begin{tabular}{|ll@{\hskip 0.05in}|@{\hskip 0.05in}r@{\hskip 0.1in}r@{\hskip 0.05in}|@{\hskip 0.05in}r@{\hskip 0.1in}r@{\hskip 0.1in}r@{\hskip 0.1in}r@{\hskip 0.05in}|}
    \hline
         &  & \multicolumn{2}{c}{\texttt{Baseline (W)}}& \multicolumn{4}{c|}{\texttt{MinEDP (W))}} \\
         
     &  & GPU & Host & DVFS GPU & DVFS Host& P\_C GPU& P\_C Host\\
    Bench & Lib &  &  &  &  &  &  \\
    \hline
    \hline
    \multirow[t]{2}{*}{dist} & jax & 254.8 ± 3.4 & 71.9 ± 0.2 & 177.4 ± 2.4 & 72.1 ± 0.1 & 182.7 ± 2.1 & 72.9 ± 0.2 \\
     & tf & 229.8 ± 6.3 & 69.5 ± 0.5 & 166.8 ± 1.2 & 70.4 ± 0.2 & 192.8 ± 6.2 & 71.9 ± 0.6 \\
    \cline{1-8}
    \multirow[t]{2}{*}{gemm} & jax & 413.5 ± 36.7 & 64.7 ± 0.2 & 257.1 ± 25.4 & 67.9 ± 0.4 & 303.3 ± 51.0 & 72.5 ± 0.5 \\
     & tf & 407.3 ± 32.0 & 68.0 ± 0.6 & 238.4 ± 10.4 & 69.4 ± 0.3 & 291.7 ± 32.6 & 71.4 ± 0.3 \\
    \cline{1-8}
    \multirow[t]{2}{*}{m\_c} & jax & 255.6 ± 7.1 & 70.0 ± 0.3 & 177.7 ± 5.4 & 70.4 ± 0.4 & 188.7 ± 2.5 & 72.7 ± 0.4 \\
     & tf & 268.4 ± 14.3 & 71.2 ± 0.8 & 193.5 ± 3.3 & 71.1 ± 0.2 & 186.9 ± 1.1 & 73.0 ± 0.2 \\
    \cline{1-8}
    \multirow[t]{2}{*}{spmv} & jax & 108.7 ± 2.2 & 69.4 ± 0.2 & 108.7 ± 2.2 & 69.4 ± 0.2 & 104.0 ± 1.1 &  72.2 ± 0.2 \\
     & tf & 188.2 ± 2.0 & 68.4 ± 0.1 & 188.2 ± 2.0 & 68.4 ± 0.1 & 173.9 ± 7.8 & 72.7 ± 0.5 \\
    \cline{1-8}
    \multirow[t]{2}{*}{stencil} & jax & 54.1 ± 6.5 & 58.5 ± 0.3 & 54.1 ± 6.5 & 58.5 ± 0.3 & 52.5 ± 8.7 & 60.2 ± 0.6 \\
     & tf & 232.8 ± 2.8 & 71.1 ± 0.2 & 168.0 ± 1.5 & 71.4 ± 0.3 & 185.7 ± 4.3 & 72.9 ± 0.3 \\
    \cline{1-8}
    \multirow[t]{2}{*}{triad} & jax & 46.8 ± 5.2 & 57.8 ± 0.5 & 46.8 ± 5.2 & 57.8 ± 0.5 & 44.2 ± 3.7 & 59.2 ± 1.0 \\
     & tf & 234.9 ± 7.2 & 66.4 ± 0.4 & 172.1 ± 11.1 & 67.7 ± 0.5 & 208.5 ± 8.0 & 70.1 ± 0.1 \\
    \cline{1-8}
    \end{tabular}
\end{table*}
\centerline

\subsubsection{Impact of the Power capping}\
\\
Looking at the lower half part of Figures \ref{fig:intel_power} and \ref{fig:nvidia_power}), we can see that {\em power capping} directly limits the power of the primary compute component. The plots show that the measured average Pkg/GPU power effectively tracks the chosen values for cap, particularly lower ones. For example, under a cap of 100 W (setting 1), the measured GPU power of A100 is near 100 W. As the cap increases, the measured power rises until a plateau near the workload's natural demand or the device's TDP  (see GEMM at 40 W on A100). Tables \ref{tab:baseline_power_EDP_intel} and \ref{tab:baseline_power_EDP_nvidia} show that the minimum EDP often occurs at intermediate or high caps; the corresponding Pkg/GPU power reflects cap level at that specific optimal point (e.g., higher power for GEMM's optimal point compared to Triad's on Intel CPU).

Similarly to DVFS, applying power capping on the GPU has little effect on the power of the RAM of the host CPU, which shows a marginal increase as the Pkg cap rises, potentially because of an increased memory traffic from a less-throttled CPU. The power of the host CPU remains consistently low in GPU power cap settings.\\

\subsubsection{Impact of the ACPI}\
\\
We focus here on AMD EPYC (bottom-half of Figure \ref{fig:amd_power}).\\
\textbf{CPU Pkg power:} With ACPI governors we get distinct Pkg power levels. \textit{'Performance'} (Setting 2) maintains high power at the level of that with high DVFS settings. \textit{'Powersave'} (Setting 1) significantly reduces the Pkg power. Intermediate governors like \textit{('ondemand', 'schedutil'} (Settings 3-6) result in average power levels, reflecting their respective DVFS policies.

\noindent\textbf{DRAM power:} DRAM power remains consistently low across all ACPI governors, thus a similar behavior as with the DVFS. The choice of the Governor does not significantly influence the DRAM power draw.\\

\subsubsection{Power profile of the frameworks}\
\\
The power draw profiles for the main compute components (Pkg/GPU) with JAX (blue line) and TF (orange line) as shown in Figures \ref{fig:intel_power}, \ref{fig:amd_power}, and \ref{fig:nvidia_power} are similar under the same settings, particularly with regard to DVFS and PowerCap. There are small, potentially benchmark-specific differences (e.g., TF's Pkg power slightly higher on AMD ACPI/DVFS plots across some benchmarks), but there is no consistent pattern of one framework always drawing significantly more compute power than the others across all scenarios.

Framework-related power differences between DRAM or Host are marginal, which suggests that potential energy gaps likely come from the primary compute component's activity.

\subsection{Result variability}

\subsubsection{General observations}\ 
\\
The measurements from our benchmarking show relatively small standard deviations, which indicates a consistent behavior from which we can say that our observations really match the effect of considered energy management techniques. \\

\subsubsection{Scenarios with higher variability}\ 
\\
In some specific scenarios, we got significant relative standard deviations:

\textbf{Low power/frequency states:} In some cases, runs at the most extreme energy-saving settings (very low frequencies through DFVS  or very tight power caps) showed significant variability in execution time. This might be due to the impact of the system when the core computation is significantly slowed down or the potential instability near hardware's operational limits. For example, the standard deviation for JAX EDP on Intel GEMM (DVFS Setting 7) are relatively higher than with other settings.

\textbf{Framework startup/JIT:} With warm-up runs considered, there might be residual JIT compilation effects (especially with JAX) or other framework initialization processes that have led to higher overhead on the first runs. Benchmarks with very short running times might be more sensitive to the aforementioned effects.

\textbf{Complex benchmarks/interactions:} Workloads involving more complex access patterns or potential OS interactions might inherently exhibit higher variations than quite regular computations like GEMM or Triad. However, no specific benchmark consistently displayed dramatically higher variability across all platforms and settings in our experiments.\\

\subsubsection{Variability between frameworks}\ 
\\
No systematic trend indicated that none of the framework between JAX and TF had more variable measurements. The variability appeared to be more closely related to the specificity of the platform, the benchmark, and the operating point (energy setting).\\


\section{Discussions}\label{sec:discuss}

\subsection{Synthesis of findings on energy management techniques}
Our empirical evaluation reveals that the effectiveness of energy management techniques is highly context-dependent. No single technique universally provides the optimal EDP across all scenarios, thus the need for a platform/workload-aware approach.

AMD Zen 3 system shows noticeable EDP improvements with DVFS for many workloads, maybe due to the design of its cache system. The dual-socket Intel system responds well to package-level Power Capping. With Nvidia A100 which has a high-TDP, we get significant efficiency gains from power capping when not applied too aggressively; this likely comes from direct 
 power draw bounding in a massively parallel architecture, where plain DVFS might not capture all power saving opportunities or might hit stability limits (as seen with JAX). These findings underscore the need for platform-aware energy management policies.

\subsection{Impact of workload characteristics}
\begin{figure*}[htbp]
    \centering
    \includegraphics[scale=0.35]{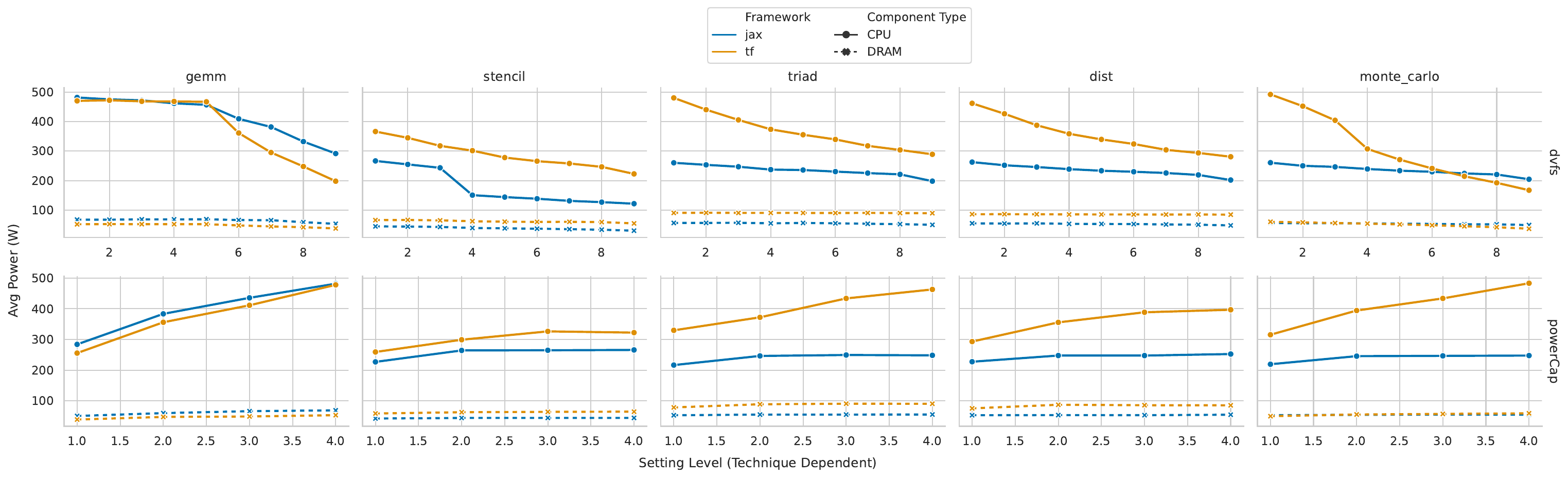}
    \caption{Power trends on Intel}
    \label{fig:intel_power}
\end{figure*}

\begin{figure*}[htbp]
    \centering
    \includegraphics[scale=0.35]{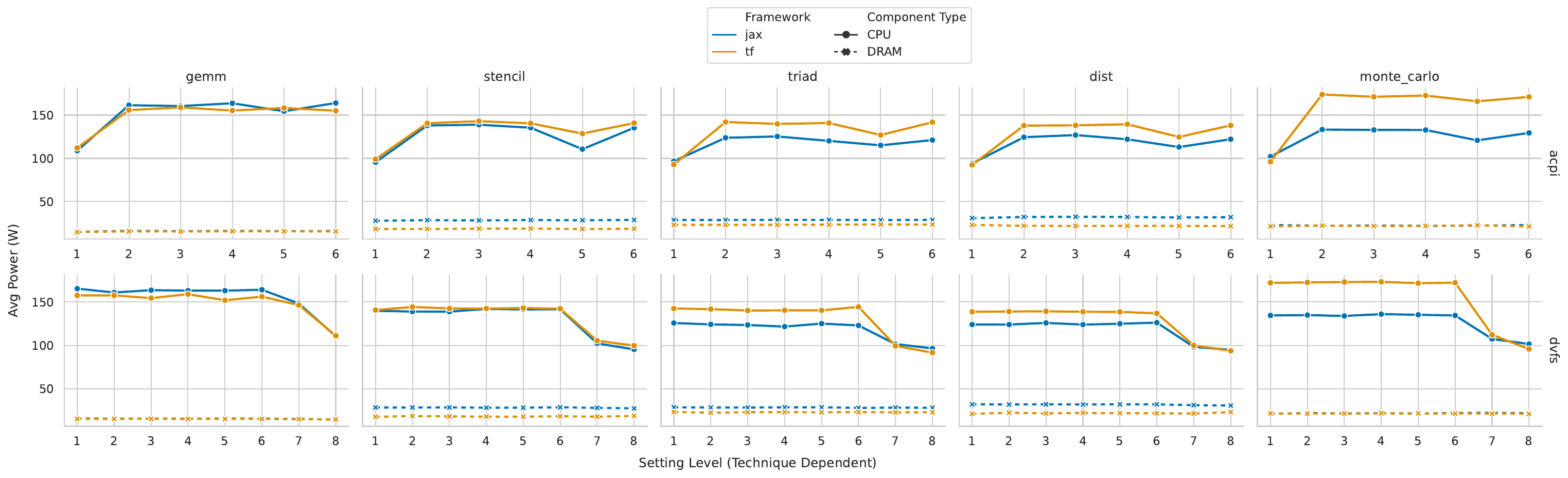}
    \caption{Power trends on AMD}
    \label{fig:amd_power}
\end{figure*}

\begin{figure*}[htbp]
    \centering
    \includegraphics[scale=0.32]{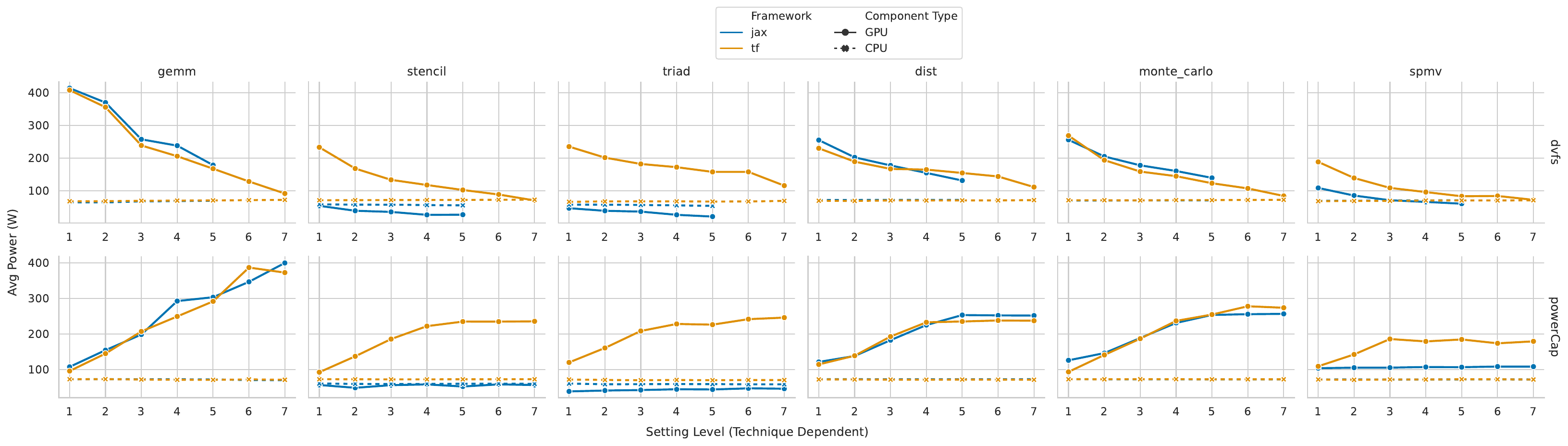}
    \caption{Power trends on Nvidia}
    \label{fig:nvidia_power}
\end{figure*}

\textbf{Compute-intensive benchmarks}, illustrated here by GEMM, have consistently exhibited the highest compute component power draw (Pkg/GPU) at baseline across all platforms (Tables \ref{tab:baseline_power_EDP_intel}, \ref{tab:baseline_power_EDP_amd} and \ref{tab:baseline_power_EDP_nvidia}). Consequently, GEMM's computation-related metrics (execution time and power consumption) seem to be highly sensitive to variations of core/GPU frequency (DVFS) or direct power limits (Power Capping), as shown in Figures \ref{fig:intel_power}, \ref{fig:amd_power} and \ref{fig:nvidia_power}. Lowering frequency or cap directly slows down the computation speed, thus leading to higher running time but with lower overall energy. However, because these workloads are compute-intensive, we might sometimes end-up with cases where the increase in time is not compensated by the energy saving, thus leading to less favorable EDP improvements compared to other workloads, particularly under severe throttling (e.g., low power caps on A100, low DVFS on CPUs).

\textbf{Memory-bound benchmarks} such as Triad, SpMV  and Dist, exhibit different characteristics. Their baseline compute component power draw was generally lower than GEMM's (Tables  \ref{tab:baseline_power_EDP_amd}, \ref{tab:baseline_power_EDP_intel} and \ref{tab:baseline_power_EDP_nvidia}), indicating less intensive FPU usage. As a result, their execution time and Pkg/GPU power were less sensitive to core/GPU frequency reductions through DVFS (top-half part of Figures \ref{fig:intel_power}, \ref{fig:amd_power} and \ref{fig:nvidia_power}). Significant performance degradation often only occurred at the very lowest frequency settings. Power capping still effectively limit Pkg/GPU power draw, but the impact on execution time was often less severe than for GEMM down to very low caps, sometimes leading to more favorable EDP reductions with moderate settings. 

\subsection{Framework runtime model and behavior}
While the primary focus of this work is on energy management techniques, the use of two prominent frameworks, JAX and TensorFlow, revealed noticeable differences in baseline {\em performance}, {\em efficiency} trends, and operational {\em robustness}. Quantitative comparisons of the measurements must consider variations in benchmark configurations (data sizes, iterations). However, analyzing the trends in power draw and the response to energy management settings offers valuable insights.

We could see that JAX and TF often induced similar power draw in the primary compute component when operating under equivalent management settings (DVFS frequency, power cap, or ACPI governor). Yet, Table \ref{tab:baseline} clearly shows significant differences in their baseline execution times for several benchmarks. This suggests that the frameworks, while capable of driving the hardware to comparable intensity levels, differ considerably in their runtime efficiency for specific tasks. We now given and described the main factors that contribute to the aforementioned observation.\\

\subsubsection{Compilation strategies}\ 
\\
JAX heavily relies on JIT compilation via {\em Accelerated Linear Algebra} (XLA), which performs specific optimizations like {\em operator fusion}. TF utilizes $tf.function$ to trace Python code and generate optimized graphs through several optimization passes that incur a different compilation overhead compared to XLA. The effectiveness of these strategies might be workload-dependent; XLA's fusion might provide significant advantages for complex computational graphs (potentially contributing to JAX's speed with Stencil or Dist), while TF's graph execution or direct kernel calls might be more optimized for some large-scale computations (like the baseline GEMM on A100).

JAX was significantly faster on Stencil than TF despite under memory pressure (according to the rematerialization warning). This strongly indicates that XLA's fusion for stencil computations ($roll$ + $add$ operations) was highly effective at reducing the total number of memory accesses and arithmetic operations, outweighing any potential slowdown due to rematerialization. TF's $@tf.function$ might not achieve the same level of fusion as the $tf.roll$ based stencil, potentially leading to more addition intermediate processing, making it slower even without rematerialization warnings. TF $tf.roll$ might internally generate copies, whereas $jax.numpy.roll$ (likely translated by XLA) might optimize this better with clever indexing/memory considerations before the computation.\\

\subsubsection{Kernel implementation \& deployment}\ 
\\
Both JAX and TF rely on back-end libraries (MKL-DNN, cuDNN, cuBLAS, etc.). However, differences in the version of the library, the way kernels are chosen and deployed, and the structure of the implementation (e.g., for sparse operations) might cause performance variations even on scenarios with similar average powers. The sparse formats used (BCOO in JAX vs. SparseTensor in TF) and their associated computational kernels are typical examples that might influence the results with SpMV.

TF's high baseline performance for large GEMM on the A100, achieved even with larger problem sizes (e.g. N = 59k vs. JAX's N = 41k) suggests highly optimized kernels or graph execution for this task, potentially enhanced by the efficiency of the $cuda\_malloc\_async$ allocator in handling all  necessary intermediate buffers.

Furthermore, the observed TF performance are without TensorRT optimizations as confirmed by runtime warnings. TF-TRT integration, which often yields significant speedups on Nvidia GPUs through specialized kernel fusion and graph optimization, wasn't used. This means that the performance gap between JAX/XLA and non-TRT/TF observed with some  benchmarks (e.g., Stencil, Dist) might narrow if TensorRT were enabled for TF. Conversely, TF's strong performance in other cases (e.g., large GEMM) was achieved even without this additional optimization layer.

\subsubsection{Runtime overheads}\ 
\\
Differences in Python execution overhead (even within $@jit$ or $@tf.function$), data marshalling between host and device, and internal scheduling could also contribute to runtime discrepancies.

\subsubsection{Memory management}\ 
\\
Memory management strategies appears to play a significant role in our observations. TF, configured to use the efficient $cuda\_malloc\_async$ allocator, demonstrated the ability to handle larger problem sizes with some benchmarks (e.g., GEMM N = 59k) compared to JAX (GEMM N = 41k). JAX required specific tuning (PREALLOCATE = false, MEM\_FRACTION = .10, ALLOCATOR = platform) to avoid out-of-memory issues with default settings on these large scenarios, suggesting the inadequacy of default approach when running at low GPU frequency (Setting 6 and Setting 7). Even with the previous tuning, JAX/XLA issued warnings related to memory pressure and rematerialization attempts with Dist and Stencil on large instances, which indicates that it was actively trading some computations for memory footprint reduction. The superior memory scalability observed with TF is likely to come from the benefit of the efficiency and lower fragmentation provided by the $cuda\_malloc\_async$ API. While our aggregate DRAM/Host power metrics did not show large discrepancies, the underlying efficiency of memory allocation significantly impacted the maximum achievable problem size and potentially had a direct influence on performance. Further study comparing different allocator choices with each framework (e.g., JAX's default BFC vs. platform, TF's BFC vs. async) would be necessary to fully isolate the aforementioned effects.\\

\subsubsection{Robustness at low power states (JAX on A100 DVFS)}\ 
\\
The most striking framework-specific observation was the inability of JAX to fully operate at the lowest DVFS clock settings on Nvidia A100 (Figs \ref{fig:nvidia_jax_edp}). TF completed all runs successfully. This two facts illustrate a significant robustness difference under specific low-power conditions. Some of the reasons for this difference include:

\textbf{Compiler sensitivity:} XLA's generated code might make assumptions about {\em minimum hardware throughput}, {\em latency}, or {\em clock frequency stability} that are violated at extremely low frequencies, leading to failures (hangs or numerical errors).

\textbf{Runtime/driver interaction:} Issues could arise in the interaction between JAX runtime, XLA backend, and CUDA driver/firmware when operating with  the GPU at low voltage/frequency.

\textbf{Kernel sensitivity:} Specific low-level kernels used by JAX/XLA might be less tolerant to the {\em lower parallelism} or {\em higher latency} associated with ultra-low clock speeds compared to those used by TF.

In essence, the choice between JAX and TF involves performance and robustness trade-offs that depend on the hardware and the nature of the workload. Although having similar power draw peaks, their respective design/implementation lead to different efficiencies and operating constraints, particularly with extreme  energy management settings.

\subsection{Technical aspects of energy-aware computing}
The results of our empirical study yield practical technical observations for users, developers, and designers involved in energy-aware computing. We present some of them.\\

\subsubsection{Sensitivity of power-related considerations}\ 
\\
Our main observation is that optimal energy management strategies are highly dependent on the considered {\em hardware platform}, the {\em computational characteristics of the workload}, and even the {\em software framework}. Just considering the default \textit{"powersave"} setting or aggressively lowering  frequency/power caps will not necessarily yield the best energy efficiency (EDP). Thus the need for a context-aware approach.\\

\subsubsection{Guidance for practitioners}\ 
\\
\indent\textbf{Platform-specific tuning:} Users should prioritize platform-dedicated techniques. Our results suggest {\em Power Capping} for Intel Xeon (good EDP often without severe performance penalty), {\em DVFS} for AMD EPYC (significant reduction in EDP with little performance slowdown) and {\em moderate Power Capping or DVFS} for Nvidia A100 GPUs (extremely low caps/frequencies  incur severe performance penalty and worsen EDP, particularly with compute-heavy tasks). ACPI governors offer simple alternatives, but less control.

\textbf{Workload characterization:} Identifying whether a workload is compute-bound or memory-bound is crucial. Compute-bound applications are generally more sensitive to frequency/power scaling and show better power reduction, but at the expense of performance. Techniques targeting core compute power will have less impact on system energy for memory-bound tasks where the power of the DRAM and the interconnects might be dominant (our results show relatively stable power of the DRAM).

\textbf{Importance of the benchmarking:} Given that power-related mechanisms are not highly predictable, seeking optimal energy efficiency should consider a suitable benchmarking with various configurations and settings in order to figure out the most impactful choices. 

\textbf{Consider framework robustness:} It is important to consider the robustness of the frameworks as some might not operate properly with particular settings. For instance, deploying JAX on Nvidia GPUs show some level of instability with very low-range DVFS settings.\\

\subsubsection{For framework developers}\
\\
\textbf{Robustness across operational settings:} Ensuring stable and correct behavior across operational settings (frequency, power states, cap values, etc...) is crucial for energy efficiency investigation. Issues with DVFS on JAX A100, for instance, illustrates possibilities for improvement.

\textbf{Energy-aware runtime/compilation:} Frameworks might benefit from energy-aware design/implementation. This could require profiling energy characteristics at compilation (like XLA or TF Graph optimization) or providing runtime APIs that allow users or schedulers to dynamically select suitable execution strategies or hardware settings based on the energy goals together with operational constraints.

\textbf{Seamless power monitoring:} Direct integration of lightweight and accurate power/energy monitoring capabilities into frameworks (maybe by extending tools like TF Profiler or JAX's utilities) could lower the barrier for users to investigate energy-related aspects.\\

\subsubsection{For hardware designers}\
\\
\noindent\textbf{Power-related elements:} Our study reinforces the importance of accurate and accessible power monitoring features of individual components (CPU Pkgs, DRAM, GPU, Host/Interconnect), which allows for better understanding of the behavior and impact of each of the main components w.r.t power-related mechanisms.

\textbf{Effectiveness of control mechanisms:} Data provides feedback on the impact of DVFS and power capping mechanisms. The mixed behavior of Power Capping on the high-TDP A100 (where aggressive caps hurt EDP) provides some hints about more skillful power management techniques or more capping strategies. The limited impact of core scaling on DRAM power highlights the need for potentially independent memory power management strategies.

Achieving significant energy savings in high-end computing systems requires a co-design approach that involves guided choices, robust/energy-aware frameworks, and computing devices that provide both effective control mechanisms and clear visibility into component-level power consumption.

\subsection{Limitations}
While this study provides valuable empirical data on energy-related mechanisms and management techniques together with frameworks behavior, we might acknowledge some limitations:

\textbf{Hardware specificity:} Our experiments were conducted on specific hardware configurations: a dual-socket Intel Xeon Platinum 8358 system, a single-socket AMD EPYC 7513 system, and an Nvidia A100 (SXM4, 40GB). Quantitative results (performance, energy, power values, optimal settings) and associated curves may differ on other-generation processors (e.g., newer EPYC Genoa/Bergamo, Intel Sapphire Rapids, Nvidia H100), socket counts, DRAM/HBM sizes/speeds/interconnects. Generalization to other hardware systems should then be done cautiously.

\textbf{Software versions:} Results are specific to the software versions used (Debian 5.10, specific JAX/TF versions, CUDA/drivers). Framework performance, compiler optimizations (XLA/TF Graph) and driver interactions evolve rapidly, newer versions might exhibit different characteristics or stability profiles.

\textbf{Measurement methodology:} Power/energy measurements rely on necessary interfaces: {\em RAPL via /sysfs} for Intel CPUs (accuracy validated but with potential limitations), {\em Linux perf events} for AMD CPUs (which can sometimes be model-dependent or less direct than RAPL), and {\em nvidia-smi} for the Nvidia GPU and Host power. The sampling interval of nvidia-smi (0.5 seconds in this study) provides average power within the indicated interval but might miss very short-lived power peaks. External high-frequency power meters could provide more precise system-level validation, but we didn't used them. Sufficient  accuracy and granularity of internal sensors are assumed to hold and account for the observed trends. The overhead of ea2p measurement tool is considered negligible.

\textbf{Lack of internal profiling:} The study focused on end-to-end application performance and system-level energy/power metrics. We did not perform deep internal profiling (e.g., {\em instruction mix analysis, cache miss rates, detailed JIT compiler logs, memory allocation tracing, fine-grained kernel execution times}). Therefore, the discussion linking observed behaviors to specific framework mechanisms ({\em compilation, memory management, kernel choice}) is somehow speculative, albeit informed by known framework characteristics. Such a detailed investigation could apply to specific behaviors like the JAX A100 low-DVFS instability.

\textbf{Need to explore other features:} We explored standard DVFS frequency steps, common ACPI governors, and a range of Power Cap limits. More advanced techniques (e.g., uncore frequency scaling, explicit C-state manipulation, combined frequency/power limits, dynamic voltage scaling if accessible) are worth investigating.

Acknowledging previously described limitations is essential to accurately explain experimental results and to guide future research directions targeting generalization and more thorough analyses.

\section{Conclusion}\label{sec:conclusion}
The primary contributions of this work are the characterization of energy/performance trade-offs across multiple relevant platforms and leading frameworks, the quantification of component-level power behavior under different management scenarios, and the identification of framework-specific operational characteristics and limitations in energy-constrained environments. These findings provide practical guidance for energy optimization purposes.

Future work should aim to broaden the scope as to include newer-generation hardware, different benchmarking with strictly controlled configurations to enable more direct framework comparisons, distributed multi-node/multi-GPU scenarios, and exploration of more advanced energy management techniques. In addition, lower-level profiling is also needed to definitively spot the root causes of performance gaps between frameworks and their robustness issues.

\section{Acknowledgements}

This research was supported by The Transition Institute 1.5 driven by École des Mines de Paris - PSL.

Experiments presented in this paper were carried out using the Grid'5000 testbed, supported by a scientific interest group hosted by Inria and including CNRS, RENATER and several Universities as well as other organizations.

\bibliographystyle{./IEEEtran}
\bibliography{./IEEEabrv,./bibliographie}

\end{document}